\documentclass[vanilla,pdf]{iucr}
\RequirePackage{graphicx}

     \journalcode{J}
\newcommand{\dd}{\textrm d}

\newcommand{\vect}[1]{\mathbf{#1}}

\usepackage{amsmath}
\usepackage{amssymb}

\begin{document}                  

\title{Solution scattering from colloidal curved plates: Barrel tiles, scrolls and spherical patches}
\shorttitle{Scattering from curved plates}


\cauthor{Doru}{Constantin}{doru.constantin@u-psud.fr}

\aff{Laboratoire de Physique des Solides, CNRS, Univ. Paris-Sud, Universit\'{e} Paris-Saclay, 91405 Orsay Cedex, \country{France}}


\maketitle                        

\begin{synopsis}
Analytical or semi-analytical expressions are derived for the orientationally-averaged scattering signal of some curved particles.
\end{synopsis}

\begin{abstract}
I provide analytical or semi-analytical expressions for the small-angle scattering of colloidal objects that can be described as curved plates. These models could help characterize a variety of inorganic or biological systems.  
\end{abstract}

\section{Introduction}

Over the last years, progress in colloidal chemistry has enabled the synthesis of a wide variety of particles, such as spheres, cubes, tetrapods, platelets with different shapes etc. \cite{Lu:2009}. In the latter category, some thinner particles do not remain flat, but rather develop a more or less pronounced curvature \cite{Levard:2010,Thill:2012}, and in some extreme cases can even roll up in scrolls \cite{Sharifi:2013,Bouet:2013,Hutter:2014}. As further examples, small building blocks (hybrid nanoparticles \cite{Park:2004} or proteins \cite{Tresset:2013}) assemble into incomplete cylindrical or spherical shells. In all of these cases, the final objects can be described as curved plates, without however reaching full spherical or cylindrical symmetry.

Due to the typical distances involved, small-angle scattering (SAS) is very well adapted to characterizing such objects. With respect to more direct techniques such as transmission electron microscopy, SAS has the advantage of intrinsically performing an average over a large number of objects (and thus of being sensitive to the sample statistics) and of being amenable to time-resolved in-situ studies, which can provide valuable information on the kinetics of various processes (e.g. of particle synthesis).

Calculating the form factor of colloidal objects is an essential pre-requisite to interpreting their SAS signal. Although analytical expressions exist for some shapes \cite{Pedersen:1997}, the procedure is rendered cumbersome by the need to integrate over all particle orientations for each value of the scattering vector. A general formalism was developed long ago \cite{Stuhrmann:1970,Svergun:1991} that expands the density distribution over the spherical harmonics and takes advantage of their integration properties to express the scattered intensity as a sum over a function basis. This method, used mainly for biomolecules in solution, considers a generic three-dimensional distribution. Alternative approaches rely on calculating the correlation function of the object in drect space and then converting to reciprocal space \cite{Glatter:1980, Ciccariello:2015}.%

In this paper, I particularize the approach of Stuhrmann and Svergun to particles that can be modeled as curved plates, with constant (or almost constant) curvature over the entire object. This strategy considerably simplifies the calculations as well as the numerical implementation of the model and is developed both for curvature along one direction (cylindrical symmetry) and for isotropic curvature (spherical symmetry). The first case applies to the rolled-up nanoplatelets discussed above, while the second one can describe, for instance, the intermediates that make up a viral capsid \cite{Tresset:2013,LawHine:2015}.

In the following, I always assume that the solutions are dilute enough for particle interactions to be negligible (the structure factor always equals 1) and that the particles are isotropically oriented, i.e. without a preferential direction. In this case, the intensity as a function of the scattering vector $I(q)$ is proportional to the orientationally averaged form factor $\left| F(q) \right|^2$ to be obtained below. The cylindrical and spherical cases will be denoted by subscripts $c$ and $s$, respectively.

\section{Methods}

\subsection{Cylindrical coordinates}

We assume that the density distribution can be separated in a contribution in the $(x,y)$ plane and along the polar axis $z$: $F_c(\vect{r}) = G(r,\phi) H(z)$. The two-dimensional Fourier transform of the in-plane part is:
\begin{equation}
\label{eq:TFcyl}
\tilde{G}(q_r,\eta) = \int_{0}^{\infty} r \dd r \int_{0}^{2 \pi} \dd \phi \exp (- i \vect{q_r} \vect{r}) G(r,\phi), 
\end{equation}
that we will expand over the basis of harmonic functions \cite{Baddour:2011}:
\begin{equation}
\label{eq:expcyl}
\tilde{G}(q_r,\eta) = \sum_{k = -\infty}^{\infty} g_k (q_r) \exp (i k \eta)
\end{equation}

When modelling the SAS signal from the particles, we will only need the rotational average (in the $(r,\phi)$ plane) of the squared modulus of $\tilde{G}(q_r,\eta)$:
\begin{equation}\label{eq:Gq}
\left\langle \left| \tilde{G}(q_r,\eta) \right|^2 \right\rangle_{\eta}  = \frac{1}{2 \pi} \int_{0}^{2 \pi} \dd \eta \left| \tilde{G}(q_r,\eta) \right|^2 = \sum_{k = -\infty}^{\infty} \left| g_k (q_r)\right| ^2
\end{equation}
\noindent which must be further combined with the Fourier transform of $H(z)$, given by:
\begin{equation}\label{eq:Hq}
\tilde{H}(q_z) = \int_{-\infty}^{\infty} \dd z \exp (- i q_z z) H(z)
\end{equation}
\noindent by an average over the polar angle $\theta$:
\begin{equation}\label{eq:Gqavg}
\left| \tilde{F}_c(q) \right|^2 = \frac{1}{2} \int_{0}^{\pi} \sin \theta \, \dd \theta \left\langle \left| \tilde{G}(q_r,\eta) \right|^2 \right\rangle_{\eta} \left| \tilde{H}(q_z) \right|^2
\end{equation}
\noindent where $q_r =\left| \vect{q_r} \right| = q \sin \theta$ and $q_z = q \cos \theta$.

\subsection{Spherical coordinates}

In spherical coordinates, the Fourier transform of a function $F_s(\vect{r})$, with $\vect{r}=(r,\theta ',\phi ')$ is:
\begin{equation}
	\tilde{F}_s(q,\theta,\phi) = \int_{\mathbb{R}^3} \dd ^3 \vect{r} \exp(-i\vect{q} \vect{r}) F_s(\vect{r})
	\label{eq:FFsph}
\end{equation}
If $F_s(\vect{r})$ can be separated into a radial part and one that only depends on the solid angle $\Omega ' = (\theta ', \phi ')$: $F_s(\vect{r}) = f(r) G(\theta ', \phi ')$ we can expand the latter over the spherical harmonics $Y_{\ell m}$, that we define as in \S~3.5 of \cite{Jackson98}. We then have $G(\theta ', \phi ') = \sum_{\ell m} c_{\ell m} Y_{\ell m} (\theta ', \phi ')$, with:
\begin{equation}
	c_{\ell m} =  \int \sin(\theta ') \, \dd \theta ' \int \dd \phi ' Y^*_{\ell m} (\theta ', \phi ')
	G(\theta ', \phi ') \label{eq:clm}
\end{equation}
and can rewrite (\ref{eq:FFsph}) as:
\begin{align}
& \tilde{F}_s(q,\theta,\phi) =  \sum_{\ell =0}^{\infty} \sum_{m=-\ell}^{\ell} c_{\ell m} \int_{0}^{\infty} r^2 \, \dd r f(r) \times\\ \nonumber
& \times \int \sin(\theta ') \, \dd \theta ' \int \dd \phi ' \exp(-i\vect{q} \vect{r}) \, Y_{\ell m} (\theta ', \phi ') 
	\label{eq:FFsphsep}
\end{align}
\noindent or, using the spherical harmonics expansion of a plane wave (\cite{Jackson98}, \S~3.6):
\begin{equation}
	\tilde{F}_s(q,\theta,\phi) = 4 \pi \sum_{\ell =0}^{\infty} \sum_{m=-\ell}^{\ell} i^{-\ell} c_{\ell m} Y_{\ell m} (\theta, \phi) b_{\ell}(q)
	\label{eq:FFsphexp}
\end{equation}
\noindent with
\begin{equation}
	b_{\ell}(q) = \int_{0}^{\infty} r^2 \, \dd r j_{\ell}(qr) f(r).
	\label{eq:flq}
\end{equation}


The rotational average of the squared modulus of $\tilde{F}_s$ is given by:
\begin{align}
\label{eq:FFsphavg}
&\left| \tilde{F}_s(q) \right|^2 = \left\langle \left| \tilde{F}_s(q,\theta, \phi) \right|^2 \right\rangle_{\Omega} \nonumber \\
&= \frac{1}{4 \pi} \int_{0}^{\pi} \sin (\theta) \, \dd \theta \int_{0}^{2 \pi} \dd \phi  \left| \tilde{F}_s(q,\theta, \phi) \right|^2 \\
&= 4 \pi \sum_{\ell =0}^{\infty} \sum_{m=-\ell}^{\ell} \left|  c_{\ell m} \right|^2 \left|  b_{\ell} (q) \right|^2 = 4 \pi \sum_{\ell =0}^{\infty} \left|  b_{\ell} (q) \right|^2 \sum_{m=-\ell}^{\ell} \left|  c_{\ell m} \right|^2 \Rightarrow \nonumber \\ 
&\left| \tilde{F}_s(q) \right|^2 = 4 \pi \sum_{\ell =0}^{\infty} c_{\ell} \left|  b_{\ell} (q) \right|^2 \label{eq:FFsphavgfinal}
\end{align}
\noindent where $c_{\ell} = \sum_{m=-\ell}^{\ell} \left|  c_{\ell m} \right|^2$ can be found either by explicitly resumming all the $m$ terms or by using the double integral:
\begin{equation}
c_{\ell} = \frac{2 \ell +1}{4 \pi} \int \dd \Omega \int \dd \Omega ' \, G(\Omega) G^*(\Omega ') \, P_{\ell} \left [ \cos(\widehat{\Omega, \Omega '}) \right ]
\label{eq:cl}
\end{equation}

\section{Results and discussion}

\subsection{Barrel tiles}\label{sec:tiles}

\begin{figure}\label{fig:Cyl_coord}
	\includegraphics[width=0.7\textwidth,angle=0]{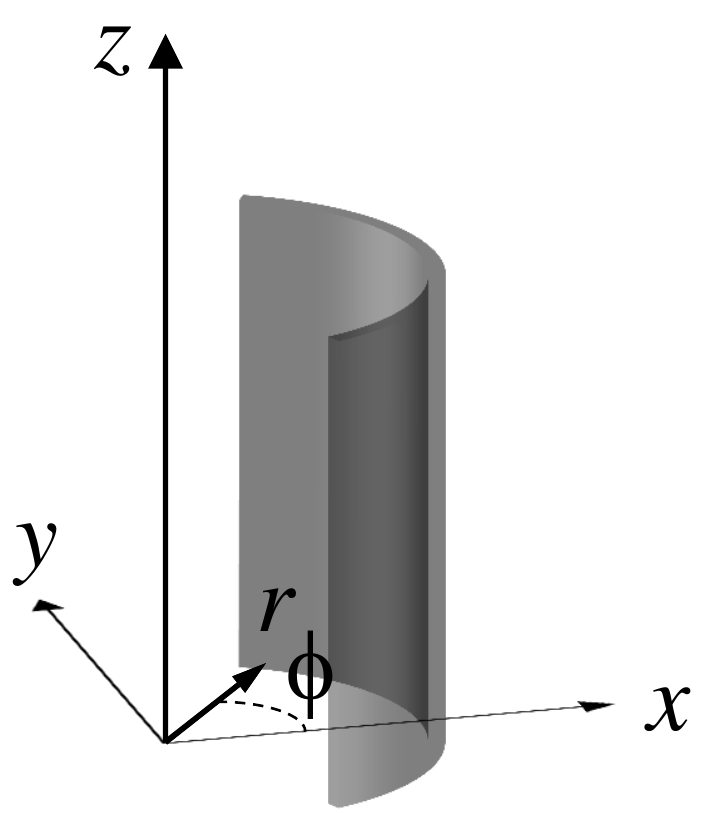}%
	\caption{Barrel tile represented in a cylindrical coordinate system.}%
\end{figure}

Figure \ref{fig:Cyl_coord} shows a cylindrically curved rectangular plate (barrel tile) in cylindrical coordinates.


For simplicity, we assume the thickness $d$ of the tile to be negligible with respect to all other length scales involved: $ G(r,\phi) = d \delta (r - R) \Theta (\phi, 0, \phi_0)$, where $R$ is the curvature radius, $\phi_0$ is the opening angle of the tile and the generalized Heaviside function $\Theta (\phi, 0, \phi_0) = 1 $ for $0 \leq \phi < \phi_0$ and 0 elsewhere. From \eqref{eq:TFcyl}:
\begin{equation}
\label{eq:TFbarrel}
\tilde{G}(q_r,\eta) = d R \int_{0}^{\phi_0} \dd \phi \exp \left[ - i q_r R \cos (\phi - \eta)\right] ) 
\end{equation}
\noindent and its coefficients $\displaystyle g_k (q_r) = \frac{1}{2 \pi} \int_{0}^{2 \pi} \dd \eta \, \tilde{G}(q_r,\eta)$ are:
\begin{equation}
\label{eq:gkbarrel}
g_k (q_r) = d R \, \phi _0 \, i^k \exp \left( \frac{i k \phi _0}{2}\right) \operatorname{sinc} \left( \frac{k \phi _0}{2}\right) J_k (q_r R)
\end{equation}
\noindent with $\operatorname{sinc}(x) = \sin(x)/x$ the cardinal sinus function, yielding for the in-plane average \eqref{eq:Gq}:
\begin{equation}\label{eq:Gqtile}
\left\langle \left| \tilde{G}(q_r,\eta) \right|^2 \right\rangle_{\eta}  = \left( d R \, \phi _0 \right)^2 \left[  J_0^2 (q_r R) + 2 \sum_{k=1}^{\infty} a_k J_k^2 (q_r R) \right] 
\end{equation}
\noindent where the coefficients $\displaystyle a_k = \operatorname{sinc}^2 \left( \frac{k \phi _0}{2}\right) $ are independent of $q_r$ and of $R$.

\subsection{Scrolls}\label{sec:scrolls}

\begin{figure}\label{fig:Scroll_diag}
	\caption{Scroll represented in a cylindrical coordinate system. The intersections of the object with the $(x,z)$ plane are highlighted.}%
	\includegraphics[width=0.7\textwidth,angle=0]{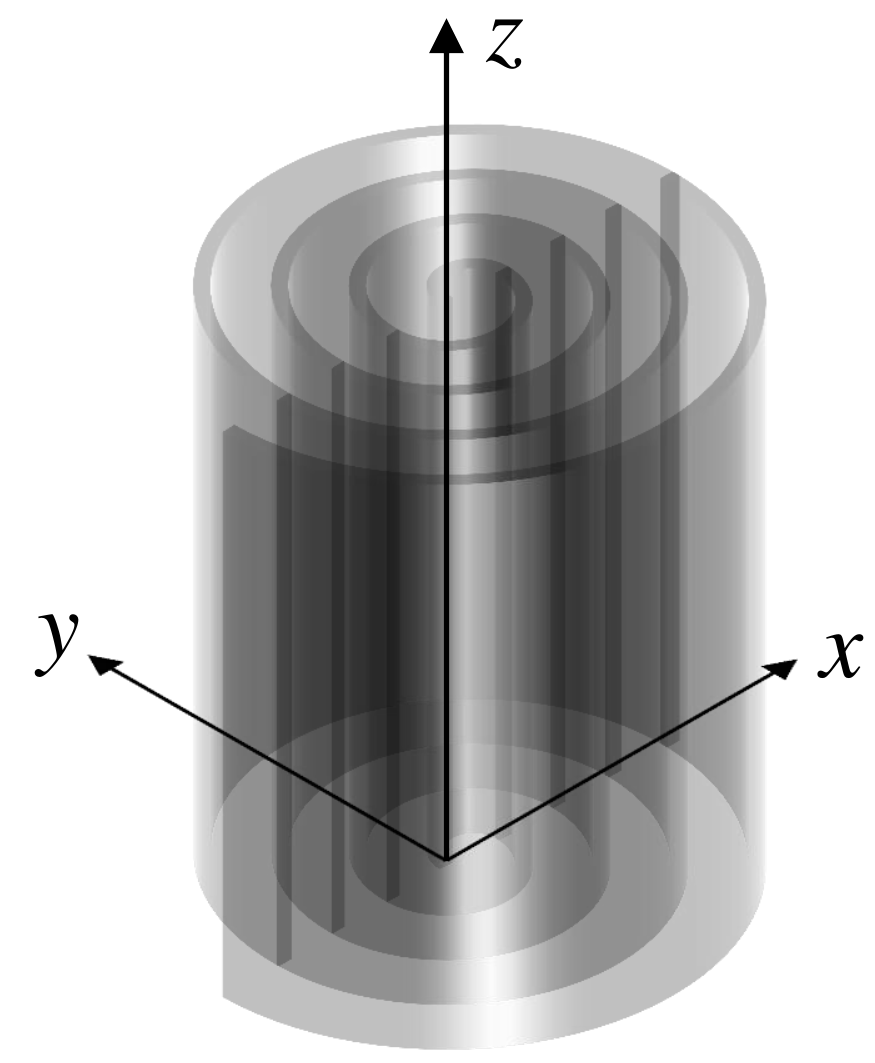}%
\end{figure}

We consider a rolled-up sheet as in Figure~\ref{fig:Scroll_diag}. Its thickness $d$ is constant and negligible with respect to the other dimensions in the problem.


We assume that the section of the object in the $(x,y)$ (or $(r,\phi)$) plane is a spiral, i.e. it can be written as $\phi(r) = f(r)$, with $f$ a monotonously increasing differentiable function. This condition implies that the object does not ``double back'' on itself, in radius or in angle.

\begin{figure}\label{fig:Scroll_geom}
	\includegraphics[width=0.9\textwidth,angle=0]{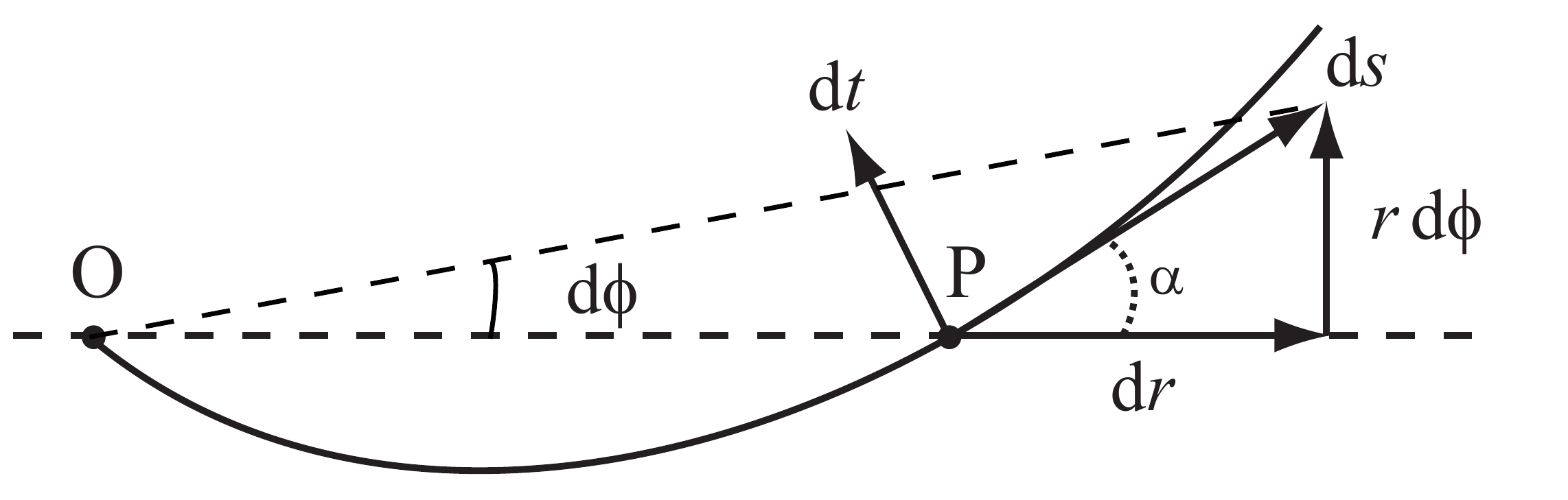}%
	\caption{Detail of the scroll section. $s$ and $t$ are the local curvilinear coordinates at point $P$ and $\alpha$ is the slope.}%
\end{figure}

In the coordinate system ($s,t$), with $s$ the local tangent and $t$ the normal to the curve (see Figure~\ref{fig:Scroll_geom}), the scroll section is given by $\delta(t)$. When converting to the polar coordinates $(r,\phi)$ one needs to account for the slope of the curve $\alpha$, with $\cos \alpha = \left[ 1+ r^2 f'(r)^2 \right] ^{-1/2}$. In particular,
\begin{equation}\label{eq:deltacurv}
\delta \left[ t(\phi)\right]  = \frac{1}{r} \sqrt{1+ r^2 f'(r)^2} \; \delta \left[ \phi - f(r)\right]
\end{equation}
The spiral is then given by:
\begin{equation}
\label{eq:Grscroll}
G(r,\phi) = \left\lbrace 
\begin{array}{cl}
 d \dfrac{\sqrt{1+ r^2 f'(r)^2}}{r} \, \delta (\phi - f(r)) & \text{for } r \leq R_0 \\
 0 & \text{for } r > R_0
\end{array}
\right. 
\end{equation}
Without the square root prefactor, Eq.~\eqref{eq:Grscroll} would describe the same geometrical shape, but with an inhomogeneous thickness (or linear density). The Fourier transform \eqref{eq:TFcyl} is:

\begin{equation}
\label{eq:TFscroll}
\tilde{G}(q_r,\eta) = d \int_{0}^{R_0} \dd r \sqrt{1+ r^2 f'(r)^2} \, \exp \left[ - i q_r \cos (f(r) - \eta) \right]  , 
\end{equation}
\noindent and its coefficients in expansion \eqref{eq:expcyl} are finite Hankel transforms of order $k$ \cite{Sneddon:1946}:
\begin{equation}
\label{eq:expscroll}
g_k(q_r) = i^k d \int_{0}^{R_0} \dd r \sqrt{1+ r^2 f'(r)^2} \, \exp \left[ - i k f(r)\right]  J_k(q_r r),
\end{equation}

\subsection{Spherical patches}\label{sec:patches}


The simplest case is that of a negligible thickness $d$: $f(r) = d \delta (r - R)$. From \eqref{eq:flq}, the coefficients are simply:
\begin{equation}
b_{\ell}(q) = d R^2 j_{\ell}(qR).
\label{eq:blq}
\end{equation}

In view of the applications, we also consider the case of a finite-thickness homogeneous shell: $f(r) = 1$ for $R_{\text{min}} \leq r < R_{\text{max}}$ (with $d = R_{\text{max}} - R_{\text{min}}$) and zero elsewhere. The components \eqref{eq:flq} are then:
\begin{equation}
b_{\ell}(q) = \int_{R_{\text{min}}}^{R_{\text{max}}} r^2 \, \dd r j_{\ell}(qr)  = \frac{1}{q^3} \int_{z_{\text{min}}}^{z_{\text{max}}} z^2 \, \dd z j_{\ell}(z).
\label{eq:flq1}
\end{equation}
\noindent with $z=qr$, $z_{\text{min}} = q R_{\text{min}}$ and $z_{\text{max}} = q R_{\text{max}}$. The integral can be expressed in terms of the regularized hypergeometric function \cite{Olver:2010}:
\begin{align}
& b_{\ell}(q) = \frac{1}{q^3} \frac{\sqrt{2}}{2^{\ell +2}} \Gamma \left( \frac{\ell + 3}{2}\right) \times \nonumber \\ 
& \left. \times z^{\ell +3} \, {_1\tilde{F}_2} \left( \frac{\ell + 3}{2}; \frac{2 \ell + 3}{2}, \frac{\ell + 5}{2}; -\frac{z^2}{4} \right) \right| _{z_{\text{min}}}^{z_{\text{max}}}
\label{eq:flq2}
\end{align}
\noindent where the vertical bar at the end indicates that the expression should be evaluated between $z_{\text{min}}$ and $z_{\text{max}}$. The components $b_{\ell}(q)$ only depend on $R_{\text{min}}$ and $R_{\text{max}}$ and can be precomputed and stored for further use if these parameters remain constant.


\section{Applications}

Let us now consider some examples. For plates (\S~\ref{sec:bending}) and spherical caps (\S~\ref{sec:sphercap}) we will compute the form factor of an object with constant volume and increasing curvature; this is useful for determining the point at which the curved-object model starts to differ from the flat-object approximation.

We will also discuss scrolls whose section is a logarithmic spiral. In this particular case, the integral \eqref{eq:expscroll} has an analytical expression in terms of regularized hypergeometric functions.

\subsection{Bending plates}\label{sec:bending}

\begin{figure}\label{fig:bending}
	\includegraphics[width=0.9\textwidth,angle=0]{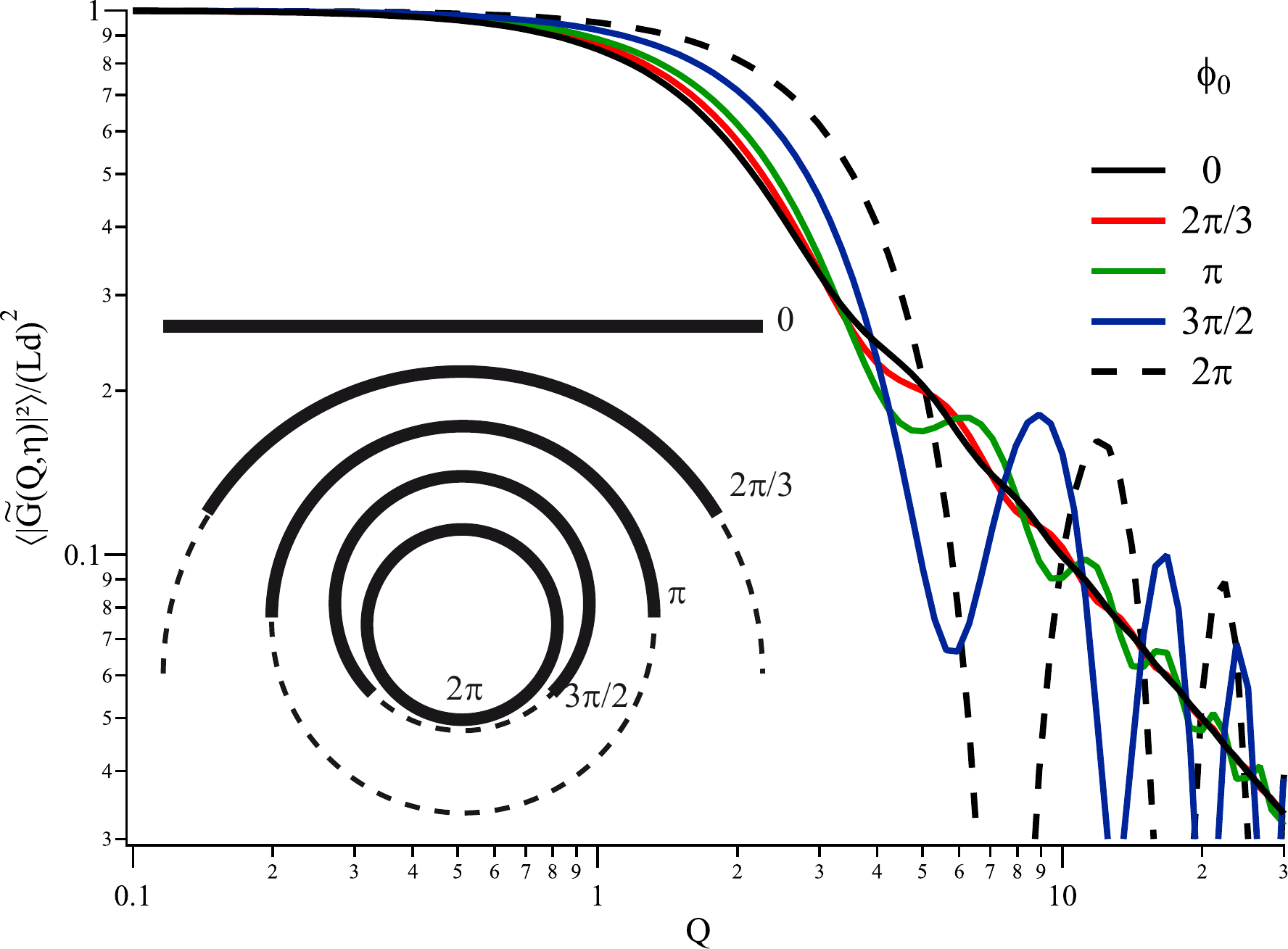}%
	\caption{Normalized in-plane average $\left\langle \left| \tilde{G}(q_r,\eta) \right|^2 \right\rangle_{\eta}/(Ld)^2$ \eqref{eq:Gqtile} as a function of the scaled scattering vector $Q=q_r L/2$ for a plate of constant length $L = R \phi_0$ and varying opening angles $\phi_0$. The shape of the plate is shown in the inset.}%
\end{figure}

We start by considering a barrel tile of constant length $L$ and with varying curvature radius $R$ and opening angle $\phi _0$, such that $R\phi _0 = L$. The in-plane average \eqref{eq:Gqtile} rescaled by its low-angle value $\left\langle \left| \tilde{G}(q_r,\eta) \right|^2 \right\rangle_{\eta}/(Ld)^2$ is shown in Figure~\ref{fig:bending} for $\phi _0$ values ranging from 0 (flat plate) to $2\pi$ (complete circle). The section of the tile is shown in the inset. The intensity in the limiting case $\phi _0 = 0$ was evaluated directly from the Fourier transform of the object, without expanding in the components \eqref{eq:gkbarrel}. 


Clearly, the effect of curvature on the scattering signal is only important for strongly curved plates, with a curvature radius $R$ smaller than about half the length $L$.

The convergence is quite fast, since the intensity no longer changes perceptibly beyond seven to ten terms.

\subsection{Logarithmic spirals}\label{sec:logspiral}

With the notations of \S~\ref{sec:scrolls},
\begin{equation}
f(r) = 2 \pi \ln \left( \frac{r}{R_c}\right), \quad \text{for } R_{\text{min}} \leq r < R_{\text{max}}
\label{eq:flog}
\end{equation}
\noindent where we need to introduce a lower limit $R_{\text{min}}$. We renormalize distances by $R_c$, defining $u= r/R_c$, $u_{\text{min}}= r/R_{\text{min}}$, $u_{\text{max}}= r/R_{\text{max}}$ and $Q = q_r R_c$. The coefficients \eqref{eq:expscroll} have a closed expression involving the regularized hypergeometric function:
\begin{align}
& g_k(Q) =  \frac{i^k \sqrt{1+(2 \pi)^2}}{2^{k+1}} \frac{\Gamma \left( \frac{k+1}{2} - k i \pi \right) }{\Gamma (k + 1) \Gamma \left( \frac{k+3}{2} - k i \pi \right)} \, Q^k \times \nonumber \\
& \left. \times u^{2k + 1 - k i \pi} \, {_1\tilde{F}_2} \left( \frac{k + 1}{2}  - k i \pi; k+1, \frac{k+3}{2}  - k i \pi; -\frac{Q^2 u^2}{4} \right) \right| _{u_{\text{min}}}^{u_{\text{max}}}
\label{eq:gklog}
\end{align}

\begin{figure}\label{fig:Logspiral}
	\includegraphics[width=0.6\textwidth,angle=0]{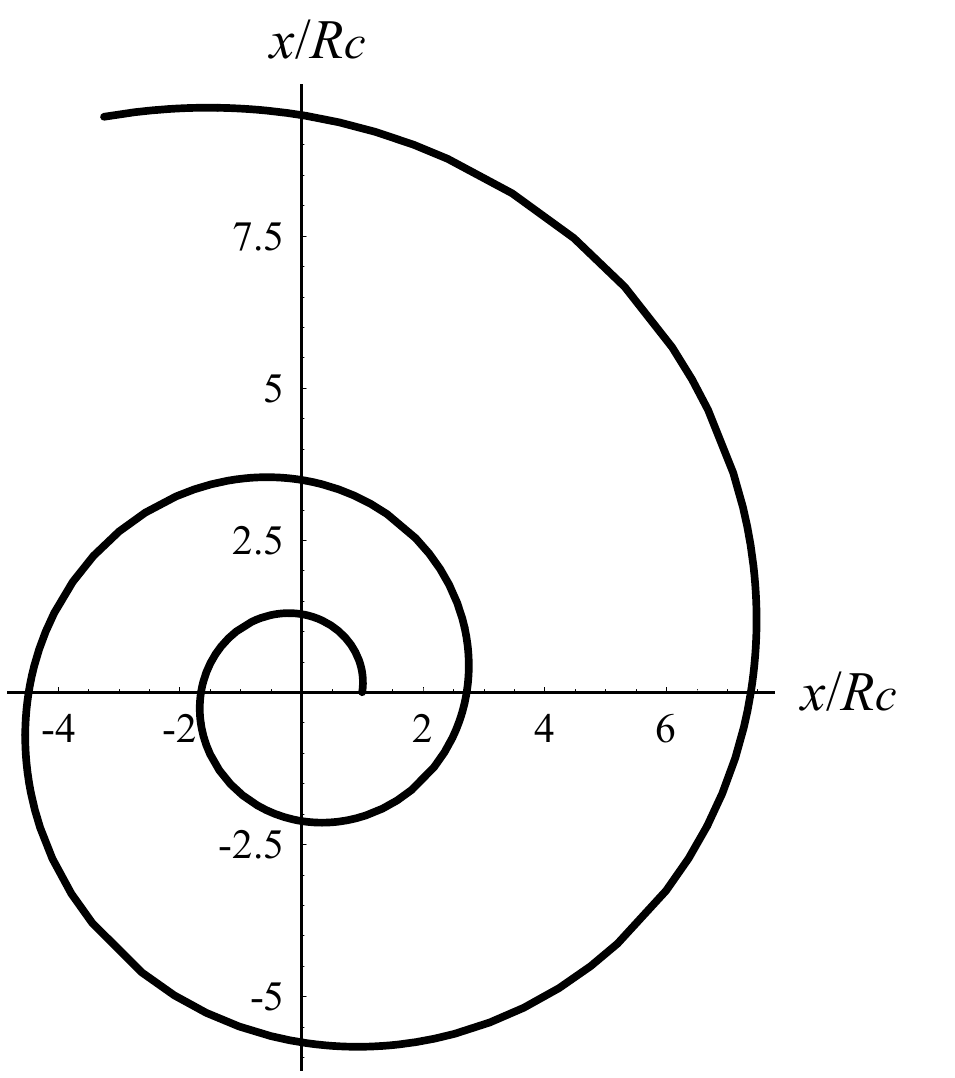}
	
	a)
	
	\includegraphics[width=0.6\textwidth,angle=0]{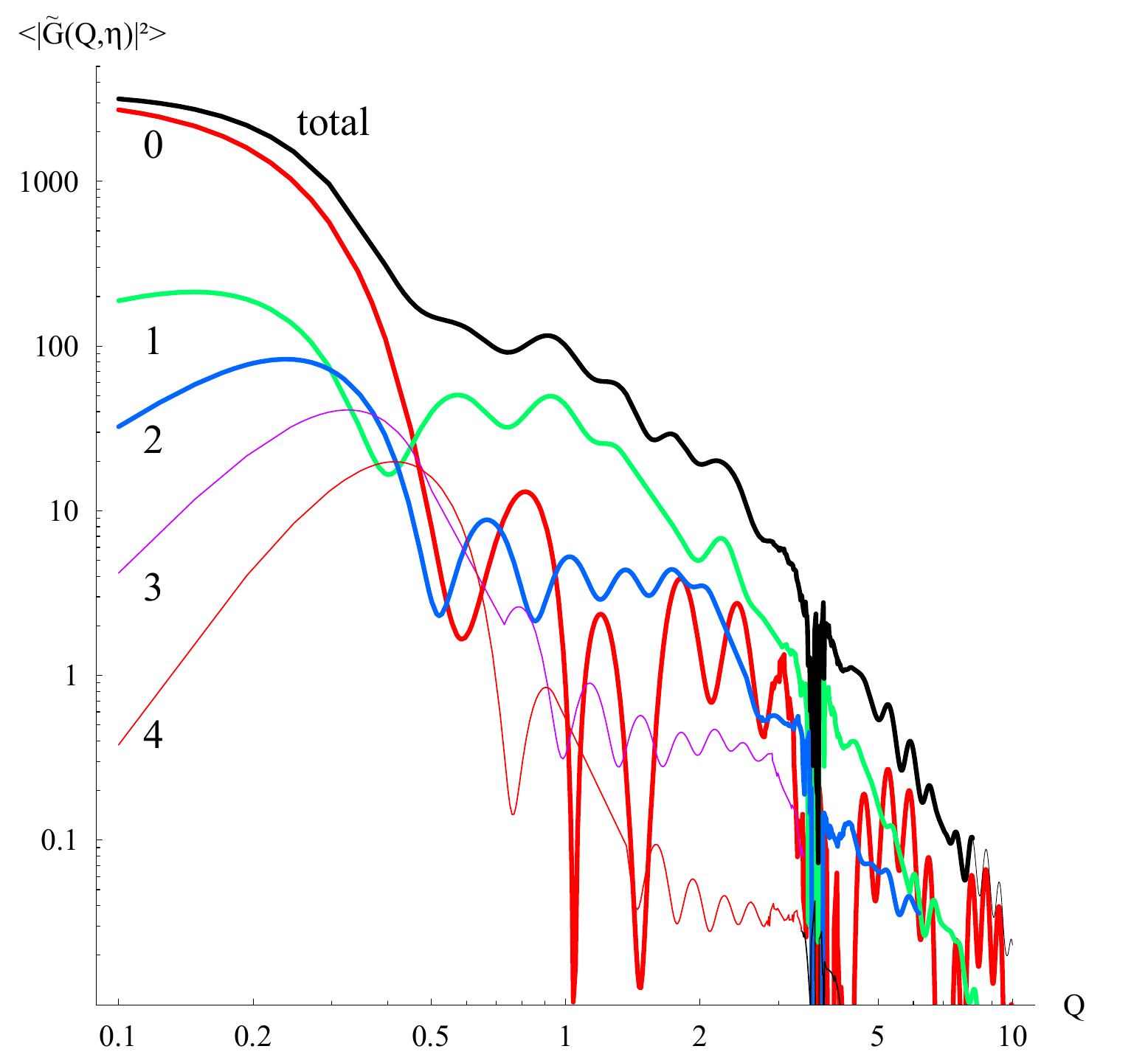}
	
	b)
	\caption{a) Logarithmic spiral \eqref{eq:flog} with $R_{\text{min}} = R_c$ and $R_{\text{max}} = 10 R_c$ b) In-plane average for the spiral in a) with $k_{\text{max}} = 4$. The individual terms are labelled and the top curve is the sum \eqref{eq:Gqlog}.}%
\end{figure}

In practice, a few ($k_{\text{max}}$ less than ten) coefficients suffice for an accurate determination of the in-plane average \eqref{eq:Gq}:
\begin{equation}\label{eq:Gqlog}
\left\langle \left| \tilde{G}(Q,\eta) \right|^2 \right\rangle _{\eta}  = \left| g_0 (Q)\right| ^2 + 2 \sum_{k = 1}^{k_{\text{max}}} \left| g_k (Q)\right| ^2
\end{equation}
\noindent as illustrated in Figure~\ref{fig:Logspiral}.

%
%
%

\subsection{Spherical caps}\label{sec:sphercap}

A particularly simple case of a spherical patch is the polar cap: $G(\theta , \phi) = 1$ for $\theta < \theta _0$ and 0 otherwise, with no dependence on $\phi$. This type of object was used to model the shape of an intermediate in the assembly of bacteriophage procapsid \cite{Tuma:2008}. The authors computed numerically the distance distributions (in real space) and obtained the scattering curves by a Fourier transform.

In our approach, the scattering signal is obtained directly in reciprocal space as a series with analytical terms since all coefficients $c_{\ell m}$ with $m \neq 0$ vanish in (\ref{eq:clm}), leaving only:
\begin{equation}
c_{\ell 0} = \left\lbrace
\begin{array}{ll}
\pi \dfrac{\left[ P_{\ell - 1}(u_0) - P_{\ell + 1}(u_0)\right]^2}{2 \ell +1} & \ell > 0\\
\pi (1-u_0)^2 & \ell = 0
\end{array}
\right. 
\label{eq:clmcap}
\end{equation}

\begin{figure}\label{fig:polarcap}
	\includegraphics[width=0.9\textwidth,angle=0]{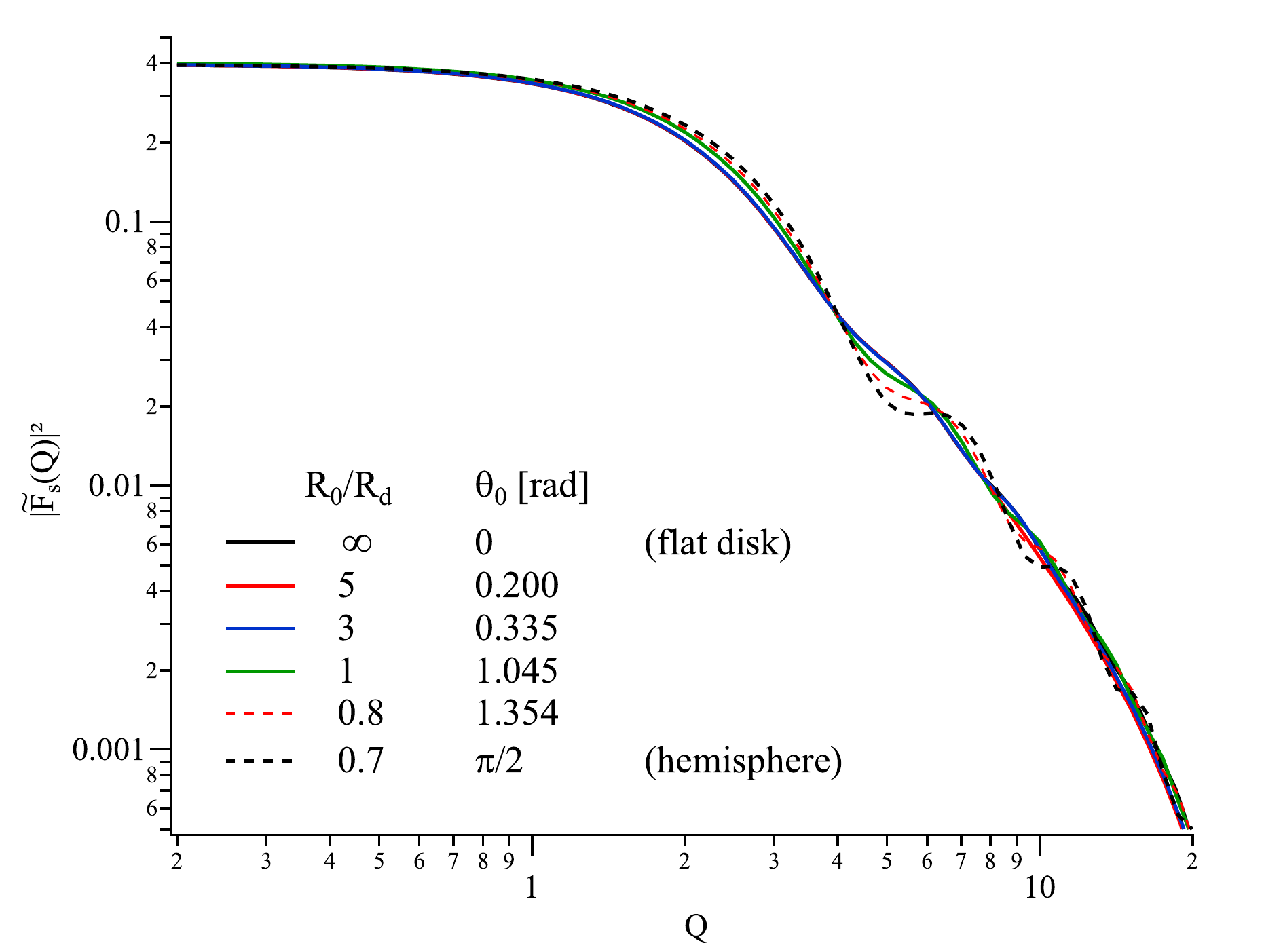}%
	\caption{Normalized average $\left\langle \left| \tilde{F}_s(q,\theta,\phi) \right|^2 \right\rangle_{\Omega}/(Ld)^2$ \eqref{eq:FFsphavg} as a function of the scaled scattering vector $Q=q R_d$ for a disk with radius $R_d$ and polar caps of equivalent volume and varying curvature radii $R_0$. For all objects, the thickness $d=R_d/5$.}%
\end{figure}

Combining equations \eqref{eq:flq2} and \eqref{eq:clmcap} it is then easy to compute the rotationally averaged scattering signal \eqref{eq:FFsphavg}. In Figure~\ref{fig:polarcap} we show this quantity for polar caps with a constant thickness $R_{\text{max}} - R_{\text{min}}$ and varying curvature radius $R_0 = (R_{\text{max}} + R_{\text{min}})/2$. The volume of the objects is chosen as equal to that of a flat disk with the same thickness $d$ and radius $R_d = 5 d$ and the scattering signal of the disk (which corresponds formally to the case $R_0 = \infty$) is also shown, but is evaluated via numerical integration rather than by the $c_{\ell 0}$ expansion above.

Identifying the volume of the disk and that of the spherical cap yields $u_0 = \cos (\theta _0)$ via:

\begin{equation}
\pi R_d^2 d = 2 \pi R_0^2 d (1-u_0) \left [ 1 + \frac{1}{12} \left(\frac{d}{R_0} \right) ^2\right ]
\end{equation}

For small curvature radii ($R_0 \sim R_d$ or smaller), the difference with respect to the flat object is visible both in the appearance of oscillations and in the reduction of the Guinier radius (the low-$Q$ plateau extends to higher $Q$ values).

In terms of convergence, ten $\ell$ terms are enough to describe the $\left| \tilde{F}_s(q) \right|^2$ \eqref{eq:FFsphavgfinal} with very good precision (down to values of about $0.01 \left| \tilde{F}_s(0) \right|^2$ ), while beyond fifteen terms there is no longer any perceptible difference in the form factor values.

\subsection{Finite thickness}\label{sec:finthick}

For simplicity, in some cases (\S~\ref{sec:tiles}, \S~\ref{sec:scrolls} and Eq.~\eqref{eq:blq} in \S~\ref{sec:patches}) I modeled the objects as infinitely thin. One can easily account for a finite thickness $d$ by multiplying the form factor with the (amplitude squared of the) Fourier transform of a solid ball with diameter $d$:
\begin{equation}
\left| R(q) \right|^2 = \left [ 3 \frac{\sin (qd/2) - (qd/2) \cos (qd/2)}{(qd/2)^3}\right ]^2
\label{eq:finite}
\end{equation}

This function is spherically symmetric, and thus unaffected by the angular averaging. It can therefore be applied as a last step, to the $\left| \tilde{F}_s(q) \right|^2 $ calculated as in Eqs.~\eqref{eq:Gqavg} or \eqref{eq:FFsphavg}.

\section{Conclusion}\label{sec:conc}

When colloidal objects can be described as curved plates, their density profile can often be written as the product of a radial profile and an angular part. In this case, the corresponding solution (orientationally averaged) form factor can be expanded into a series, each term being representing the contribution of a given harmonic degree. 

This separation greatly reduces the computational requirements, since the number of numerical integrals required for each value of the scattering value $q$ is halved (in cylindrical coordinates) or eliminated (in spherical coordinates). Moreover, it renders the result easier to understand by connecting the $q$-dependence of the form factor to the angular density profile.

For scrolls, although only one space direction is decoupled from the angular dependence, the particular shape of the section affords a similar expansion, as demonstrated for logarithmic spirals.

\appendix
\ack{Acknowledgements}

I acknowledge fruitful discussions with Didier Law-Hine and funding from the Agence Nationale pour la Recherche under contract MEMINT (2012-BS04-0023).


\begin{figure}\label{fig:Sph_Coord}
	\includegraphics[width=0.7\textwidth,angle=0]{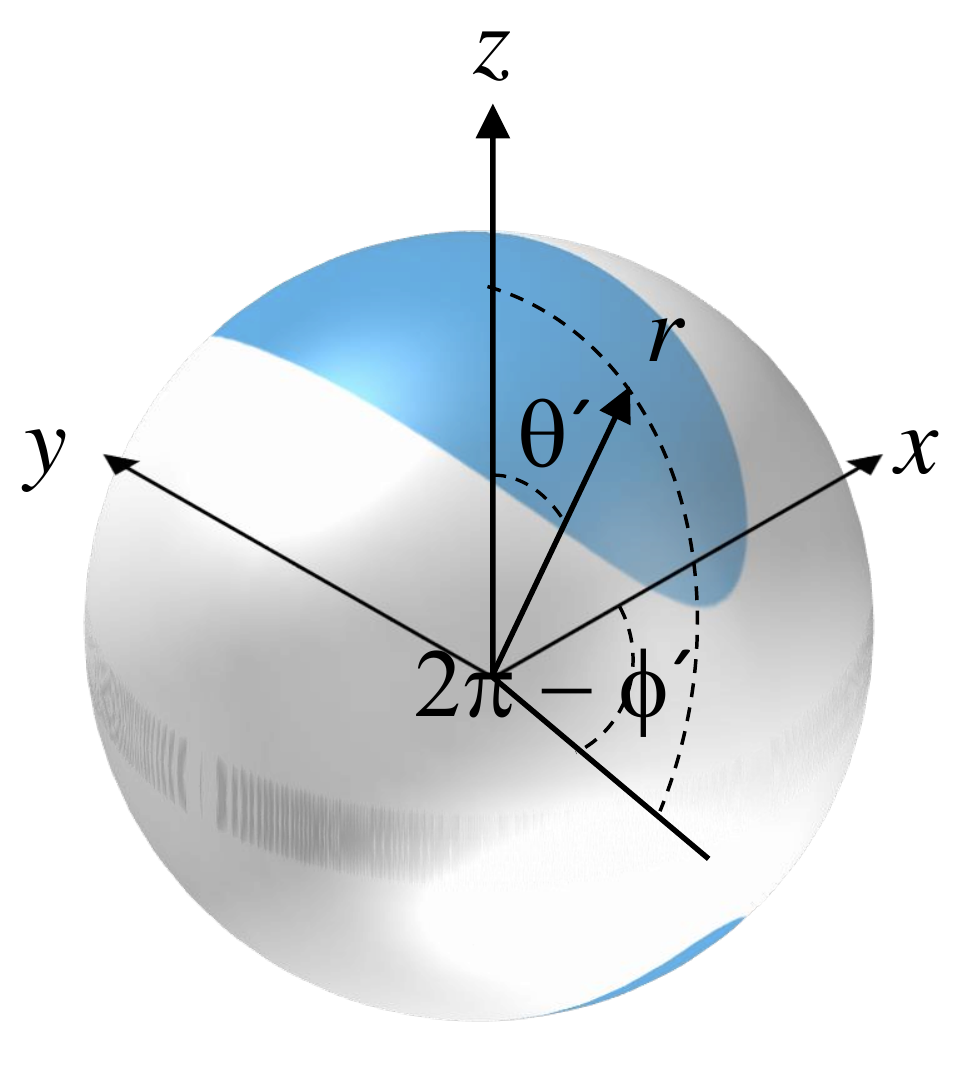}%
	\caption{Spherical patch represented in a spherical coordinate system.}%
\end{figure}

\referencelist[curved]
\end{document}